%% file: main.tex
\documentclass[12pt,a4paper]{article}
\pdfoutput=1
\usepackage{amsthm,bm,amsmath,amssymb,epsfig,multicol,multirow}
\usepackage{natbib} 
\usepackage{graphics,rotating,graphicx}
\usepackage{array,multirow,caption}
\usepackage{amssymb}
\usepackage{amsthm}
\usepackage{bm,amsmath,multicol}
\usepackage{booktabs}
\usepackage[colorlinks,bookmarksopen,bookmarksnumbered,citecolor=red,urlcolor=red]{hyperref}
\usepackage{pdflscape}
\usepackage{adjustbox}
\usepackage{tablefootnote}

\DeclareMathOperator*{\argmin}{arg\,min}

\theoremstyle{definition}

\begin{document}

\title{An efficient estimator of the parameters of the Generalized Lambda Distribution}

\author{Dilanka S. Dedduwakumara \textsuperscript{a}, Luke A. Prendergast \textsuperscript{a} and Robert G. Staudte \textsuperscript{a} \\ \\ \textsuperscript{a}Department of Mathematics and Statistics, La Trobe University}

\maketitle


\vspace{0.5cm}

\noindent\textbf{ABSTRACT.  Estimation of the four generalized lambda distribution parameters is not straightforward, and available estimators that perform best have large computation times.  In this paper, we introduce a simple two-step estimator of the parameters that is comparatively very quick to compute and performs well when compared with other methods.  This computational efficiency makes the use of bootstrapping to obtain interval estimators for the parameters possible.  Simulations are used to assess the performance of the new estimators and applications to several data sets are included.}

\vspace{0.5cm}


\section{Introduction}

The generalized lambda distribution \citep[GLD,][]{ramberg1974approximate} is a flexible, four parameter distribution that can approximate a large variety of distributions of varying shapes.  Due to this flexibility, the GLD has become a popular distribution to model data in many fields, including economics and finance \citep[e.g.][]{pfaff2016financial}.  Given that it has four parameters, location, scale and two shape parameters, estimation is not a trivial task, and estimation methods continue to attract attention in the literature.     
\cite{doi:10.1080/25742558.2019.1602929} introduced a method for choosing optimal parameters for generalized distributions to approximate other distributions.  The method uses the probability density quantile function \citep[pdQ,][]{staudte2017} to first find the optimal shape parameters and is computationally quick, simple to implement and often outperforms other methods.   Motivated by this, we introduce estimators of the GLD shape parameters arising from the estimated pdQ.  The estimators, which compare favourably to other methods, is computationally efficient, providing estimates in just a fraction of the time required for other methods that are good estimators of the parameters.  This efficiency makes obtaining interval estimators for the GLD parameters via bootstrapping possible, which is a considerable advantage given the lack of intervals in the literature.

We begin by providing important definitions and notations in Section 2 before reviewing several popular estimators already available in Section 3.  The new estimator is introduced in Section 4 and performance is assessed by simulations in Section 5.  Several data applications are given in Section 6 before we provide concluding remarks in Section 7.

\section{Definitions}

Throughout, let $Q(u)=F^{-1}(u)$ denote the quantile function associated with distribution function $F$ for $u\in (0,\ 1)$, and let $f = F'$ denote the probability density function where $f(x)> 0$ for all $x$ in its domain. 

\subsection{Probability density quantile functions}

The \textit{quantile density function} \citep{parzen1979}, also called the \textit{sparsity index} by \cite{tukey1965}, is denoted $q(u)=Q'(u)=1/f\left[Q(u)\right]$. The quantile density function is mainly used in non-parametric modeling and inference. \cite{parzen1979} called the reciprocal of this quantile density function the \textit{density quantile function} which we denote here by $f_Q(u) = f[Q(u)].$ In a recent paper, \cite{staudte2017} introduced the \textit{probability density quantile} (pdQ) denoted $f^*_Q(u)=f_Q(u)/\kappa$ where $\kappa = E[f_Q(U)]$ and $U\sim \text{Unif}(0,1)$. This pdQ function is free from location and scale and therefore can easily be used to examine shape behaviours of a distribution; see also \cite{staudte2018} who provide other insights for the pdQ. Moreover, it is defined on the finite domain $[0,1]$ for all lattice distributions and continuous distributions having square-integrable densities. 
 
\subsection{Empirical pdQ for continuous distributions}

Let $X_1,\ldots,X_n$ denote a random sample of size $n$ from $F$ and let $X_{(1)}\leq X_{(2)} \leq \ldots \leq X_{(n)}$ denote the ordered sample.  In this section we introduce the empirical pdQ earlier discussed by \cite{staudte2017}.  For $k_b(·) = k(\cdot/b)/b$ denoting a kernel function and $b$ a bandwidth, we start by estimating $q(u)$ using the quantile density estimator
\begin{equation}\label{eqn:qnhat}
\hat{q}_n(u)=\sum_{i=1}^{n} X_{(i)}\bigg\{k_b\bigg(u-\frac{(i-1)}{n}\bigg)-k_b\bigg(u-\frac{i}{n}\bigg)\bigg\}
\end{equation}
which consists of a linear combination of order statistics. This kernel density estimator has been studied extensively, e.g. see \cite{jones1992estimating}, \cite{falk1986estimation} and \cite{welsh1988asymptotically} for some notable works.  The choice of bandwidth is important and we choose our bandwidth to be $b(u)=(15/n)^{1/5}\{q(u)/q^{''}(u)\}^{2/5}$ since it minimizes the asymptotic mean squared error of $\hat{q}(u)$. \cite{prendergast2016exploiting} call $q(u)/q^{''}(u)$ the quantile optimality ratio (QOR) and estimate it to obtain a suitable bandwidth.


By using a discrete set of $u$s defined by $\{u_j=(j-1/2)/J\}^J_{j=1}$ and for some integer $J$, the empirical pdQ can be defined as 
\begin{equation}\label{eqn:fQ*hat}
\hat{f}_{Q}^*(u_j)=\frac{1}{\hat{\kappa}\hat{q}_n(u_j)} \quad \quad \text{where} \quad \hat{\kappa}=\frac{1}{J}\sum_{j=1}^{J}\frac{1}{\hat{q}_n(u_j)}.
\end{equation}

\subsection{The Generalized Lambda Distribution}

The generalized lambda distribution (GLD) is a very flexible distribution that can approximate, or is equal to exactly, many other distributions for appropriately chosen parameters.   Although several definitions for the GLD are available, we prefer the parameterization by \cite{freimer1988} since it is defined for all parameter value choices, with the exception that the scale parameter must be positive.  The distribution is defined in terms of its quantile function which is 
\begin{equation}\label{eqn:Q}
Q(u)=\lambda_1+\frac{1}{\lambda_2}\left[\frac{u^{\lambda_3}-1}{\lambda_3} - \frac{(1-u)^{\lambda_4}-1}{\lambda_4}\right]
\end{equation}
where $\lambda_1$ is a location parameter, $\lambda_2>0$ an inverse scale parameter and $\lambda_3, \lambda_4$ are shape parameters.    It is easy to see that the quantile density function for the GLD is
$q(u)=Q'(u)=\lambda_2^{-1}\left[u^{\lambda_3-1} + (1-u)^{\lambda_4-1}\right]$ so that the density quantile function is
\begin{equation}\label{fQ:GLD}
f_Q(u)=\frac{\lambda_2}{u^{\lambda_3-1} + (1-u)^{\lambda_4-1}}.
\end{equation}
In general, no closed-form solution for the integral $\kappa =\int^1_0 f_Q(u)du$ exists, but it can be evaluated computationally and quite efficiently since integration occurs only between the finite bounds zero and one. 

\section{Existing Methods}\label{Methods}

Having four parameters, GLD estimation is not straightforward, and several estimation methods are available in the literature. For comparison with the pdQ method, we consider the more common estimation methods that are believed to provide good estimates of the true parameter values.  These methods are available in either or both of the \textit{gld} \citep{Kingt2016gld} and  \textit{bda} packages \citep{wang2015bda} in the R statistical software \citep{R}. We briefly describe those estimation methods and an overview with more technical details can be found in \cite{dean2013improved}.    

The trimmed L-moments method (TL) is discussed in \cite{asquith2007moments} and \cite{dean2013improved} and is a robust version of the matching L-moments method \citep{karvanen2002adaptive}. L-moments are estimated by linear combinations of order statistics and are related to conventional moments. Trimmed L-moments (see: \cite{elamir2003trimmed} for more details) are generalizations of these L-moments which applies zero weight to extreme observations. GLD parameters are estimated through this method by minimizing the difference between the sample trimmed L-moments of the data and the trimmed L-moments of the fitted GLD distribution. 

The percentile matching method (PM) equates a selected number of empirical percentiles with their GLD counterparts to obtain a set of non-linear equations which are then solved to obtain the optimal GLD parameters.  The PM method is discussed by \cite{karian1999fitting} and \cite{tarsitano2005estimation}. Further, \cite{karian2003comparison}  demonstrate the superiority of the PM method compared to the method of matching moments and the method of L-moments. 

Maximizing the likelihood (ML) to obtain the GLD parameters has also been a popular method and is considered by \cite{su2007fitting} and \cite{su2007numerical}. First, initial estimated parameter values using the method of moments or the PM method are chosen and used as a starting point to seek the values that maximize the numerical log likelihood.

There are instances where the Maximum Likelihood method fails as the support depends upon the parameters to be estimated. In such a scenario, the Maximum product of spacings (MPS) was introduced by  \cite{cheng1983estimating,ranneby1984maximum} and used by \cite{chalabi2012flexible} to estimate GLD parameters. Here spacings refer to the differences between the cumulative distribution function at neighbouring data points and the parameters are estimated by maximizing the geometric mean of these spacings.  

Rather than the spacing between transformed data points as in MPS, \cite{titterington1985comment} suggested spacing between transformed, adjacently averaged data points.  This approach is called the Titterington Method (TM), and more details with the GLD estimation can be found in \cite{dean2013improved}.

\cite{owen1988starship} introduces the Starship Method (SM), a computer-intensive method that focuses on the fit to the base distribution of the inversely transformed data. \cite{king1999starship} developed this concept to be used for generalized lambda distribution where for a given data set the distribution function of GLD is obtained numerically, and the parameter values are chosen such that it minimizes the goodness-of-fit to the uniform distribution. One major drawback of this method is its slow computation time, especially with large sample sizes.   

The method of distributional least absolutes (DLA) obtains the optimal GLD parameters by minimizing the sum of absolute deviations between the order statistics and the corresponding medians. See \cite{dean2013improved} for more details.

\section{Point and interval estimators using the pdQ}

Obtaining optimal GLD parameters for a specific distribution using the pdQ method has been discussed in \cite{doi:10.1080/25742558.2019.1602929} and we adopt this idea to model empirical data as detailed in the steps below. Throughout this section the estimated $p^{th}$ quantile, the empirical pdQ of the data and the corresponding GLD pdQ are denoted by $\widehat{x}_p$, $\hat{f}_{Q}^*$ and $f_{Q}^*$ respectively.  

\subsection{Point estimators}

\subsubsection*{Step 1: Estimating the shape parameters}

Since the pdQ is free of location and scale, we can use it to focus only on estimating the two shape parameters, $\lambda_3$ and $\lambda_4$.  Using a discrete set of probability values given as $\{u_j=(j-1/2)/J\}^J_{j=1}$, the estimated shape parameters are those that minimize the sum of squared distances between the empircal pfdQ and GLD pdQ.  That is,
\begin{equation}\label{eqn:obj}
(\widehat{\lambda}_3,\widehat{\lambda}_4)=\argmin_{\lambda_3,\lambda_4} \sum^J_{j=1}\left[\hat{f}_{Q}^*(u_j)-f_Q^*(u_j;\lambda_3,\lambda_4)\right]^2.
\end{equation}

To find the shape parameters that minimize the sum of squared distances, it is simple to use a computational optimizer.  We use the R function \textit{nlminb} in \textit{stats} package for this, which we find to be both simple to use and quick to compute.  As starting values for the optimizer
we chose to apply the objective function in \eqref{eqn:obj} to a grid of values for $\lambda_3$ and $\lambda_4$ consisting of each paired combination of $\lambda_3,\lambda_4$ from $−0.9, −0.5, −0.1, 0, 0.1, 0.2, 0.4, 0.8, 1, 1.5$ and choose those values that result in the smallest value. This grid of values are used in \cite{dean2013improved} and considered as the default choice in \textit{gld} \citep{Kingt2016gld} package for other GLD estimation methods. This covers a wide span of the $\lambda_3,\lambda_4$ values and also we can expect uniformity between methods for the comparisons to follow. Our simulations, to be summarized in the next section, reveal that $J=50$ is more than adequate to obtain good estimates for moderate to larger sample sizes and there is little value in increasing $J$ to be more than this.  Therefore, in what follows, we use $J=50$ for sample sizes greater than 200.  We found that $J=25$ worked well for smaller sample sizes.

\subsubsection*{Step 2: Estimating the location and scale parameters}

Given the estimated shape parameters from Step 1, we then match sample quartiles $\widehat{x}_p$, using $p \in \{0.25,0.5,0.75\}$, to their theoretical GLD counterparts to generate the linear equations
\begin{equation}\label{eqn:Q}
\widehat{x}_p=\lambda_1+\frac{1}{\lambda_2} c(p,\widehat{\lambda}_3,\widehat{\lambda}_4) 
\end{equation}
where,
\begin{equation*}
c(p,\widehat{\lambda}_3,\widehat{\lambda}_4)= \left[\frac{p^{\widehat{\lambda}_3}-1}{\widehat{\lambda}_3} - \frac{(1-p)^{\widehat{\lambda}_4}-1}{\widehat{\lambda}_4}\right]
\end{equation*}

By solving the the system of three linear equations from \eqref{eqn:Q} using $p=0.25,0.5$ and $0.75$, we obtain our estimated inverse scale and location parameters as 
\begin{equation}
\widehat{\lambda}_2=\frac{c(0.75,\widehat{\lambda}_3,\widehat{\lambda}_4)-c(0.25,\widehat{\lambda}_3,\widehat{\lambda}_4)}{\widehat{x}_{0.75}-\widehat{x}_{0.25}},\;\;\hat{\lambda}_1=\widehat{x}_{0.5}-\frac{c(0.5,\widehat{\lambda}_3,\widehat{\lambda}_4)}{\widehat{\lambda}_2}
\end{equation}    
respectively.

\subsection{Interval estimators}

In order to calculate confidence interval estimates, we consider two commonly used bootstrap procedures; the percentile method \citep{efron1994introduction} and the bias-corrected and accelerated (BCa) bootstrap intervals \citep{efron1987better}. As the pdQ method requires little computational time, the computational cost to calculate the bootstrap confidence intervals is not prohibitive. 

The percentile bootstrap method uses the upper and lower $\alpha/2$ percentiles of the GLD sample estimators as the interval bounds, whereas the BCa method also estimates a bias-corrected parameter and an acceleration parameter. The bias-correction parameter is related to the proportion of bootstrap estimates that are less than the observed statistic, and the acceleration parameter is proportional to the skewness of the bootstrap distribution. Unlike the percentile method, the BCa method corrects for both bias and skewness of the estimators. 

In the following section to come, we look at the confidence intervals for the location ($\lambda_1$) and the skewness ($\lambda_3-\lambda_4$) of the underlying GLD distribution by using these bootstrap interval estimators. Note that the difference $\lambda_3-\lambda_4$ may be of interest since $\lambda_3-\lambda_4=0$ indicates that the distribution is symmetric.

\section{Simulation results}

In this section, we compare our pdQ estimation method with the existing methods presented in Section \ref{Methods}.  To do so, we summarize the results for 500 simulated data sets by reporting the mean standard error and the mean absolute bias for estimates of the four GLD parameters compared to their true values. We adopt four representative FMKL GLD settings used in \cite{corlu2016estimating} to evaluate the performances of each method for these different shapes of GLD.  The shapes of the probability density functions for these GLD settings are shown below in Figure \ref{fig1}.

 \begin{figure}[h!t]
\centering
\includegraphics[height=14cm,width=\textwidth]{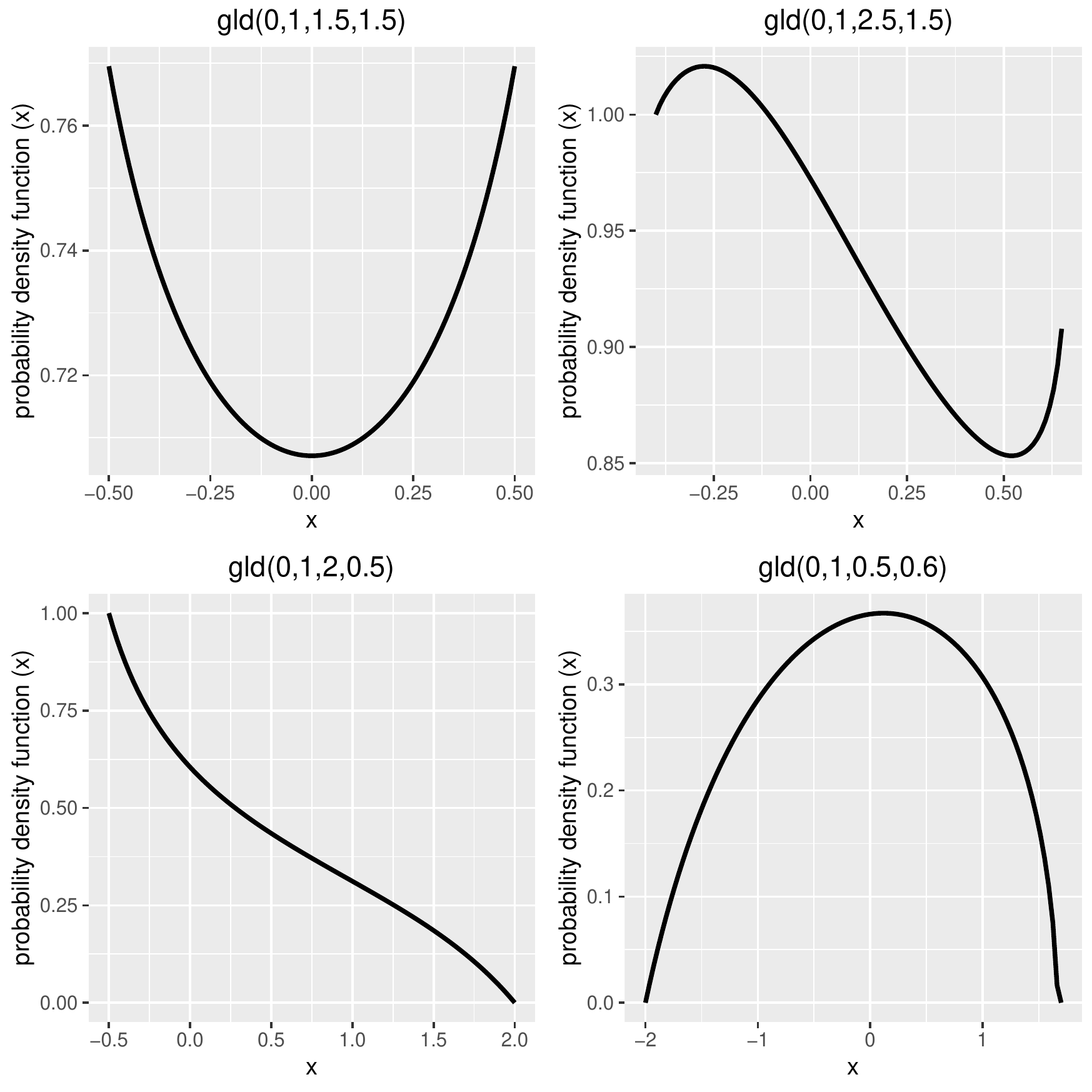}
\caption{Probability density functions of four representative  GLDs.}
\label{fig1}
\end{figure}

 Results are evaluated for a range of sample sizes:  100, 250, 500 and 1000. In the pdQ method, for sample size 100, we take $J=25$, and for other sample sizes, we use $J=50$. 

\begin{landscape}
\input{table1}
\end{landscape}

\begin{landscape}
\input{table2}
\end{landscape}

\begin{landscape}
\input{table3}
\end{landscape}

\begin{landscape}
\input{table4}
\end{landscape}

As can be seen in Table \ref{tab1}, the pdQ method shows comparatively low  mean squared error and absolute bias for the $\lambda_1 = 0,\lambda_2 =1,\lambda_3 =1.5,\lambda_4 =1.5$ setting for all the sample sizes. The MPS and TM methods also perform well, being second best or third best in most instances. Similar results can be found in Table \ref{tab2} for $\lambda_1 =0,\lambda_2 =1,\lambda_3 =2.5,\lambda_4 =1.5$ as well, where the pdQ method has the minimum MSE and bias values in almost all cases. The DLA and MPS methods also perform well for this setting. Although the pdQ is not the best method always for the last two GLD settings, $\lambda_1 =0,\lambda_2 =1,\lambda_3 =2,\lambda_4 =0.5$ and $\lambda_1 =0,\lambda_2 =1,\lambda_3 =0.5,\lambda_4 =0.6$, it produces competitive results compared to the other methods. The MPS and TM methods again provide very good results for these two settings, displaying low MSE and absolute bias values (Table \ref{tab3} and \ref{tab4}).  However, their good performance comes with much greater computational cost which we will consider below.

\begin{figure}[h!]
\centering
\includegraphics[height=17cm,width=\textwidth]{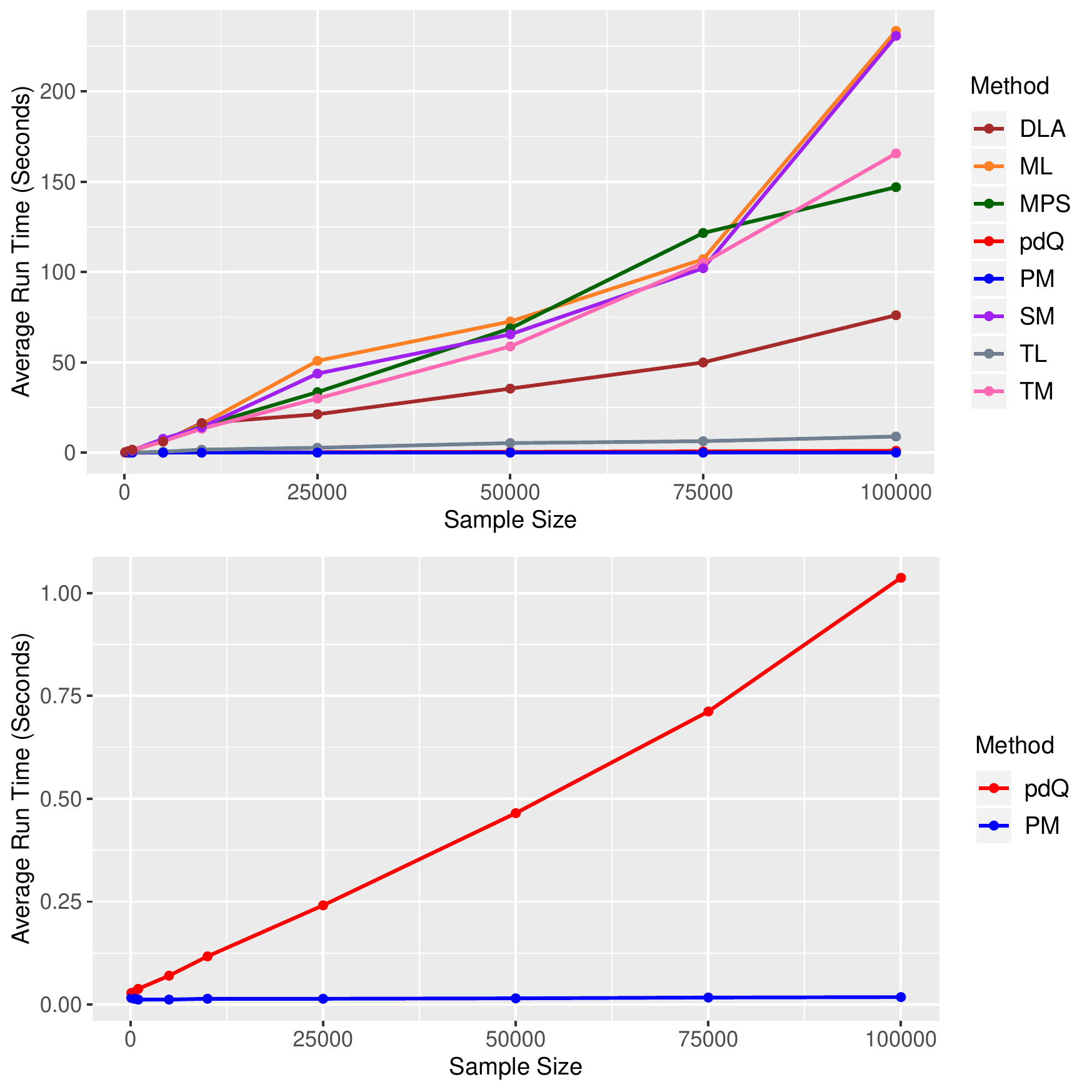}
\caption{This figure shows the variations of the average running times for each method as the sample size increases. The values are calculated using 100 iterations}
\label{fig2}
\end{figure}

In Figure \ref{fig2}, we depict the computational time for each method as the sample size increases. We consider 100 replications of data from GLD(0,1,0.5,0.6) for each sample size and calculate the average computational time for each estimation method. The run times were measured on a computer with an Intel(R) Core(TM) i7-6700 processor running at 3.40GHz using 32 GB of RAM, running Windows version 10.   The graph shows that the PM and pdQ methods have much lower computational times compared to all other methods. The PM takes just a few milliseconds to estimate the parameters from a sample even as large as 100,000 observations.  However, the PM method was typically not a good estimator when compared to the pdQ, MPS and TM methods. On the other hand, the pdQ method requires about a second to compute even for 100,000 observations. As expected, SM and ML are the slowest among all the methods, spending almost 4 minutes each to calculate the above results. The TM method has a computation time of around 2.5 minutes for this setting.  Such large computation times make bootstrapping onerous. 

\input{table5.tex}

 We now consider the performance of bootstrap intervals for some common distributions. The closest GLD parameters for these considered distributions are adopted from \cite{doi:10.1080/25742558.2019.1602929} and are presented in Table 5. Here one sample is generated from the above closest GLD parameters and bootstrapped from those observations. 
\input{table6.tex}


\input{table7.tex}

In Table \ref{tab6} and Table \ref{tab7} we provide the coverage probability results for bootstrap confidence intervals for both location ($\lambda_1$) and the difference of shape parameters ($\lambda_3-\lambda_4$), where $\lambda_3-\lambda_4$ is an indication of the skewness of the GLD distribution. We only selected the TM, PM methods together with PdQ methods for the comparison as they are the close competitors in performance wise and computational time wise.

From Table \ref{tab6}, all the methods display coverage probabilities close to the nominal value for the location parameter. The TM method has the narrower width, with the pdQ method the next narrowest. We found that larger sample sizes are required to obtain reliable intervals for the difference in shape parameters.  The TM method is not considered here as BCa bootstrap confidence intervals need a higher number of re-samples and when the sample size increases to 1000, and it takes roughly 33.98 minutes to obtain a single confidence interval for TM. The pdQ and PM methods, on the other hand, take only 47.65 seconds and 32.02 seconds subsequently to provide a single confidence interval for $n=1000$. Compared to the PM method, the pdQ method shows better coverage in the vicinity of the nominal level for skewness with a narrower width. So it can be concluded that the pdQ is favourable compared to these other two methods.

\section{Applications}
\subsection{Application 1: Total income data of private households of Spain in 1980 }

\begin{figure}[h]
\centering
\includegraphics[height=12cm,width=13cm]{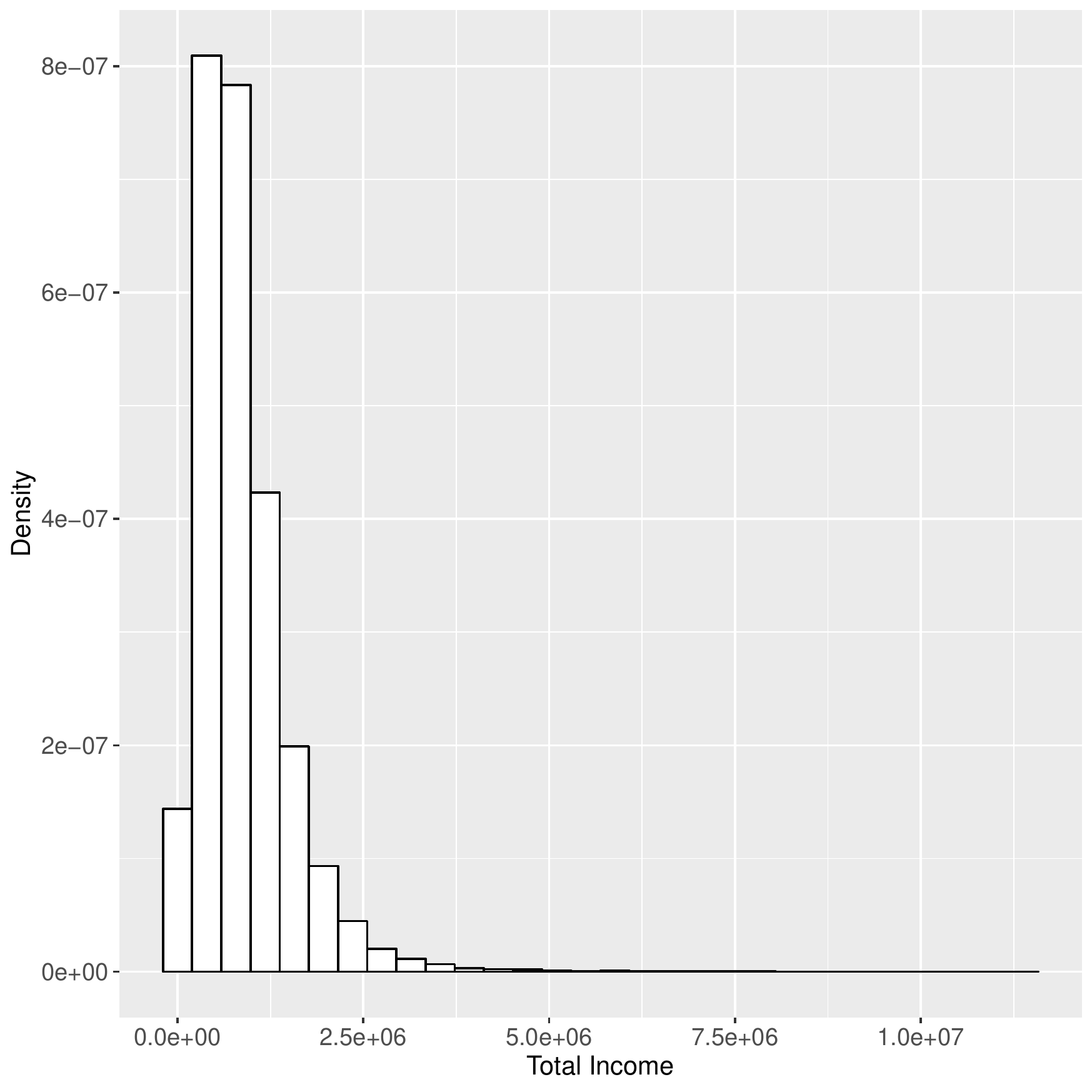}
\caption{Total income of Spanish households}
\label{fig3}
\end{figure}

In this example, we consider total income data of Spanish households (Figure \ref{fig3}), which is from the 1980 Spanish Family Expenditure Survey (FES) described in \cite{alonso1994encuesta}. This data set consists of 23,972 observations, and total income is recorded with household characteristics and expenditure on several categories. This data set is readily available in the \textit{Ecdat} package \citep{croissant2016Ecdat} under the name `BudgetFood'. 

\input{table8}

\begin{figure}
\centering
\includegraphics[height=16cm,width=15cm]{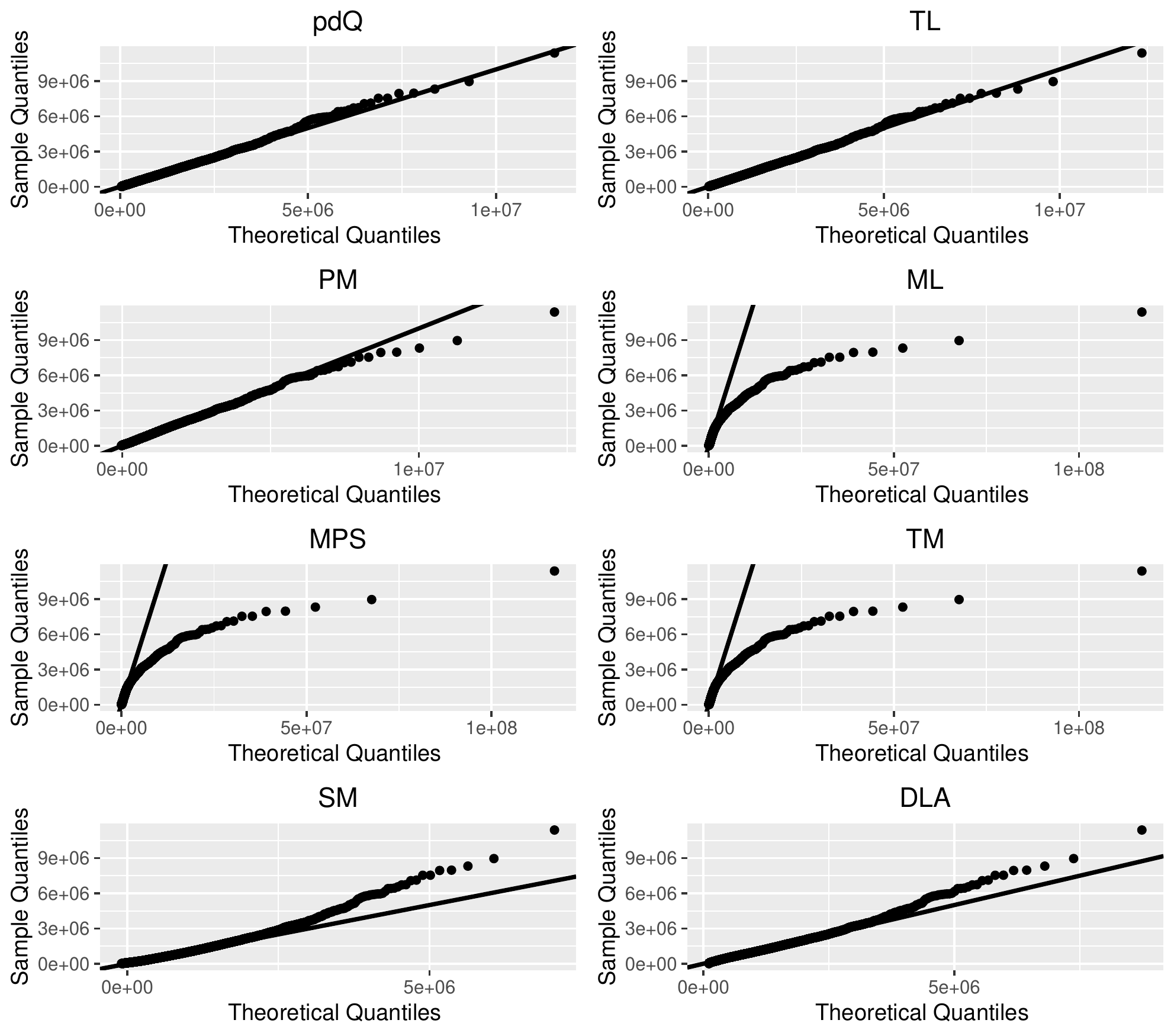}
\caption{Quantile plots for Spanish households data }
\label{fig4}
\end{figure}

In Table \ref{tab8}, we assess the goodness of fit of each method to this data by using the Kolmogorov Smirnov test statistic. As can be seen, both TL-moments and pdQ method provide the best fit to the data. Further, this is elaborated in Figure \ref{fig4} quantile plots where, unlike the other methods, estimated quantiles from the pdQ method and TL-moments method are clearly aligned with the sample quantiles.  

\input{table9}

In Table \ref{tab9}, we provide the running times for parameter estimation by each method. The PM and pdQ methods are clearly the fastest. The pdQ is about five times faster than the TL-moments method and 91 times faster than SM for this particular data set. This further suggests the suitability of the pdQ method with large sample sizes.   

\subsection{Application 2: Twin Study data }

We now consider a much smaller data set known as the Indiana Twin Study data, which consists of the birth weights of a set of 123 twins. The data originally comes from the PhD thesis of Dr Cynthia Moore, Department of Medical and Molecular Genetics, Indiana University School of Medicine and also has been used in \cite{karian2003comparison} and \cite{karian2000fitting}, for GLD parameter estimation.  

Table \ref{tab10} presents the Kolmogorov Smirnov test statistic for the GLD fit to the data by each method and the corresponding p-value for the test. According to the p-values, all the methods suggest a better fit to the data, but in particular, the SM method, TL-moments method and pdQ method are the best fits, as they display smaller KS statistics compared to the other methods.

\input{table10}

\subsection{Application 3: Earnings data }

We now look at obtaining 95\% bootstrap confidence intervals for the difference of the location parameter for two groups and also confidence intervals for the skewness. The data set we consider here is also available in the \textit{Ecdat} package \citep{croissant2016Ecdat} and is named `CPSch3'. The data set includes the hourly earnings of males ($n=5956$) and females ($n=5174$) in the US from 1992 to 1998. 

\input{table11.tex}
\input{table12.tex}
\input{table13.tex}

In Tables \ref{tab11},\ref{tab12} and \ref{tab13} we display the bootstrap confidence intervals calculated using three methods and the time efficiency of each method. For percentile bootstrap intervals, we used 500 bootstrap samples, but for BCa confidence intervals, the number of re-samples has to be much higher than the sample size. Therefore, we used 15,000 bootstrap re-samples to obtain the BCa confidence intervals as the sample sizes here are large.

All the methods provide similar results for the confidence intervals with narrower width while declaring there is a difference in the location for both males and females in earnings.  Additionally, there is some skewness present in both the male and female earnings distributions. When the width of the confidence interval is considered, pdQ has a narrower width after the TM method and outperforms the PM method. When the time efficiency is considered the pdQ and PM take only 0.945 minutes and 0.405 minutes, respectively, to obtain the percentile confidence interval whereas the TM method spends around 3.5 hours to obtain the same results. For the BCa confidence intervals, TM method is much more inefficient in this case as it takes roughly 4.5 days to obtain the BCa confidence interval for this data whereas the pdQ method and PM method only takes 59.17 minutes and 45.03 minutes respectively.

\section{Discussion}

In this paper we have introduced a simple two-step estimation method for the GLD parameters.  The first step uses the empirical probability density quantile function to find estimates of the shape parameters.  The second step can then simply obtain estimates for the location and inverse scale.  How simulations show that the method performs very well, often beating existing methods.  An additional advantage of the approach is the very small computation time compared to the best of the methods.  This small computation time makes bootstrapping possible which is important given the lack of interval estimators available for the GLD parameters.  For the more time consuming estimation methods, bootstrap intervals can take hours, even days to compute, as our examples show.

While we focused on the GLD distribution in this paper, the method can be used for other distributions too.  An obvious extension would be to consider other generalized distributions, like the generalized Beta distribution, since estimation for such distributions is not straightforward.

\bibliographystyle{authordate4}
\bibliography{ref}

\end{document}

%% file: table1.tex
\begin{table}[h]

  \centering
  
    \caption{This table shows a comparison of the mean standard error and the absolute bias of the GLD parameters represented by $\lambda_1$,$\lambda_2$,$\lambda_3$ and $\lambda_4$ between several fitting methods. The considered methods are Method of Probability Density Quantiles  (pdQ), Method of TL-moments (TL), Percentile Matching Method (PM), Numerical Maximum Likelihood (ML), Maximum Product of Spacings (MPS), Titterington's Method (TM), Starship Method (SM), and Method of Distributional Least Absolutes (DLA). These values are calculated for 500 fitting results for actual parameters $\lambda_1=0$, $\lambda_2=1$, $\lambda_3=1.5$ and $\lambda_4=1.5$.  }
\label{tab1}
\hspace*{-0.5cm}
\begin{adjustbox}{max width=1.3\textwidth}
\begin{tabular}{ccccccccccc}
\toprule
$n$ & Est & pdQ & TL & PM & ML & MPS & TM & SM & DLA\\
\midrule
100 & $\lambda_1$ & 0.013  (0.002) & 0.012  (0.005) & 0.016  (0.005) & 0.008  (0.004) & 0.011*  (0.004) & 0.011*  (0.004) & 0.012  (0.005) & 0.014  (0.005)\\
 & $\lambda_2$ & 0.158  (0.245) & 0.314  (0.368) & 0.703  (0.531) & 0.202  (0.366) & 0.136*  (0.240) & 0.142  (0.246) & 0.280  (0.333) & 0.205  (0.263)\\
 & $\lambda_3$ & 0.176  (0.219) & 0.259  (0.333) & 0.404  (0.382) & 0.184  (0.329) & 0.175*  (0.210) & 0.176  (0.204) & 0.241  (0.281) & 0.226  (0.219)\\
 & $\lambda_4$ & 0.154*  (0.220) & 0.252  (0.311) & 0.390  (0.362) & 0.184  (0.314) & 0.171  (0.196) & 0.177  (0.188) & 0.237  (0.260) & 0.210  (0.201)\\

\midrule
250 & $\lambda_1$ & 0.006  (0.006) & 0.005  (0.006) & 0.008  (0.009) & 0.004*  (0.006) & 0.005  (0.003) & 0.005  (0.003) & 0.005  (0.004) & 0.007  (0.006)\\
 & $\lambda_2$ & 0.070*  (0.155) & 0.192  (0.301) & 0.331  (0.379) & 0.184  (0.353) & 0.086  (0.197) & 0.087  (0.198) & 0.141  (0.249) & 0.122  (0.191)\\
 & $\lambda_3$ & 0.094*  (0.150) & 0.185  (0.283) & 0.261  (0.300) & 0.171  (0.324) & 0.110  (0.186) & 0.104  (0.184) & 0.145  (0.232) & 0.156  (0.151)\\
 & $\lambda_4$ & 0.094*  (0.168) & 0.185  (0.309) & 0.268  (0.343) & 0.182  (0.345) & 0.118  (0.199) & 0.120  (0.195) & 0.156  (0.245) & 0.155  (0.182)\\

\midrule
500 & $\lambda_1$ & 0.003  (0.002) & 0.003  (0.002) & 0.005  (0.004) & 0.002  (0.004) & 0.003  (0.001) & 0.002*  (0.000) & 0.003  (0.002) & 0.004  (0.003)\\
 & $\lambda_2$ & 0.056*  (0.146) & 0.135  (0.266) & 0.206  (0.308) & 0.181  (0.365) & 0.070  (0.195) & 0.069  (0.195) & 0.094  (0.209) & 0.091  (0.163)\\
 & $\lambda_3$ & 0.080*  (0.155) & 0.146  (0.275) & 0.193  (0.280) & 0.176  (0.350) & 0.092  (0.208) & 0.089  (0.207) & 0.110  (0.213) & 0.114  (0.148)\\
 & $\lambda_4$ & 0.074*  (0.161) & 0.148  (0.283) & 0.201  (0.301) & 0.186  (0.364) & 0.090  (0.211) & 0.091  (0.208) & 0.111  (0.221) & 0.119  (0.161)\\

\midrule

1000 & $\lambda_1$ & 0.002  (0.002) & 0.001*  (0.001) & 0.002  (0.002) & 0.001*  (0.003) & 0.001*  (0.001) & 0.001*  (0.000) & 0.002  (0.000) & 0.002  (0.001)\\
 & $\lambda_2$ & 0.037*  (0.095) & 0.094  (0.230) & 0.119  (0.231) & 0.174  (0.362) & 0.065  (0.196) & 0.066  (0.198) & 0.061  (0.175) & 0.058  (0.119)\\
 & $\lambda_3$ & 0.058*  (0.108) & 0.115  (0.261) & 0.136  (0.244) & 0.174  (0.358) & 0.084  (0.221) & 0.085  (0.220) & 0.082  (0.196) & 0.083  (0.121)\\
 & $\lambda_4$ & 0.061*  (0.102) & 0.112  (0.255) & 0.129  (0.233) & 0.182  (0.367) & 0.086  (0.216) & 0.085  (0.221) & 0.079  (0.194) & 0.082  (0.116)\\
\bottomrule
\multicolumn{2}{l}{\textsuperscript{*}{Lowest MSE}}
\end{tabular}
\end{adjustbox}
\end{table}

%% file: table2.tex
\begin{table}[h]

  \centering
  
    \caption{This table shows a comparison of the mean standard error and the absolute bias of the GLD parameters represented by $\lambda_1$,$\lambda_2$,$\lambda_3$ and $\lambda_4$ between several fitting methods. The considered methods are Method of Probability Density Quantiles(pdQ), Method of TL-moments (TL), Percentile Matching Method (PM,) Numerical Maximum Likelihood (ML), Maximum Product of Spacings (MPS), Titterington's Method (TM), Starship Method (SM), and Method of Distributional Least Absolutes (DLA). These values are calculated for 500 fitting results for actual parameters $\lambda_1=0$, $\lambda_2=1$, $\lambda_3=2.5$ and $\lambda_4=1.5$  }
\label{tab2}
\hspace*{-0.5cm}
\begin{adjustbox}{max width=1.3\textwidth}
\begin{tabular}{ccccccccccc}
\toprule
$n$ & Est & pdQ & TL & PM & ML & MPS & TM & SM & DLA\\
\midrule
100 & $\lambda_1$ & 0.014  (0.081) & 0.013*  (0.084) & 0.015  (0.079) & 0.013*  (0.094) & 0.014  (0.084) & 0.013*  (0.083) & 0.013*  (0.083) & 0.014  (0.079)\\
 & $\lambda_2$ & 0.972  (0.820) & 1.513  (1.049) & 2.607  (1.264) & 0.922  (0.898) & 0.844*  (0.774) & 0.896  (0.791) & 1.443  (1.002) & 1.156  (0.871)\\
 & $\lambda_3$ & 1.665  (1.212) & 2.088  (1.382) & 2.211  (1.374) & 1.909  (1.355) & 1.689  (1.184) & 1.662*  (1.177) & 1.980  (1.313) & 1.726  (1.207)\\
 & $\lambda_4$ & 0.299*  (0.436) & 0.490  (0.606) & 0.664  (0.644) & 0.330  (0.517) & 0.314  (0.448) & 0.311  (0.449) & 0.468  (0.560) & 0.400  (0.492)\\
 
\midrule
250 & $\lambda_1$ & 0.010*  (0.083) & 0.010*  (0.088) & 0.011  (0.082) & 0.011  (0.097) & 0.011  (0.088) & 0.011  (0.088) & 0.010*  (0.086) & 0.010*  (0.077)\\
 & $\lambda_2$ & 0.622*  (0.688) & 1.142  (0.972) & 1.493  (1.019) & 0.837  (0.881) & 0.648  (0.719) & 0.669  (0.733) & 0.982  (0.888) & 0.808  (0.746)\\
 & $\lambda_3$ & 1.530*  (1.170) & 2.063  (1.399) & 1.987  (1.328) & 1.974  (1.387) & 1.656  (1.214) & 1.671  (1.220) & 1.900  (1.320) & 1.577  (1.147)\\
 & $\lambda_4$ & 0.219*  (0.395) & 0.427  (0.601) & 0.499  (0.594) & 0.317  (0.535) & 0.249  (0.448) & 0.255  (0.452) & 0.364  (0.550) & 0.319  (0.454)\\
\midrule

500 & $\lambda_1$ & 0.008*  (0.079) & 0.009  (0.087) & 0.008*  (0.076) & 0.010  (0.097) & 0.010  (0.090) & 0.010  (0.091) & 0.009  (0.087) & 0.008*  (0.074)\\
 & $\lambda_2$ & 0.569*  (0.671) & 1.050  (0.965) & 1.151  (0.929) & 0.873  (0.915) & 0.617  (0.728) & 0.629  (0.735) & 0.895  (0.883) & 0.731  (0.742)\\
 & $\lambda_3$ & 1.496*  (1.162) & 2.065  (1.415) & 1.841  (1.297) & 2.058  (1.426) & 1.699  (1.254) & 1.702  (1.261) & 1.930  (1.356) & 1.571  (1.158)\\
 & $\lambda_4$ & 0.215*  (0.405) & 0.422  (0.618) & 0.446  (0.592) & 0.338  (0.567) & 0.249  (0.464) & 0.251  (0.465) & 0.359  (0.563) & 0.309  (0.484)\\

\midrule
1000 & $\lambda_1$ & 0.008  (0.083) & 0.009  (0.091) & 0.007*  (0.078) & 0.011  (0.100) & 0.010  (0.097) & 0.010  (0.097) & 0.009  (0.093) & 0.007*  (0.074)\\
 & $\lambda_2$ & 0.623*  (0.722) & 0.998  (0.975) & 0.904  (0.858) & 0.926  (0.957) & 0.675  (0.791) & 0.685  (0.798) & 0.848  (0.892) & 0.674  (0.716)\\
 & $\lambda_3$ & 1.659  (1.231) & 2.140  (1.456) & 1.763  (1.285) & 2.169  (1.471) & 1.870  (1.346) & 1.877  (1.347) & 2.008  (1.402) & 1.548*  (1.135)\\
 & $\lambda_4$ & 0.239*  (0.440) & 0.414  (0.632) & 0.385  (0.568) & 0.356  (0.593) & 0.270  (0.502) & 0.273  (0.507) & 0.346  (0.574) & 0.288  (0.471)\\
\bottomrule
\multicolumn{2}{l}{\textsuperscript{*}{Lowest MSE}}
\end{tabular}
\end{adjustbox}
\end{table}

%% file: table3.tex
\begin{table}[h]

  \centering
  
    \caption{This table shows a comparison of the mean standard error and the absolute bias of the GLD parameters represented by $\lambda_1$,$\lambda_2$,$\lambda_3$ and $\lambda_4$ between several fitting methods. The considered methods are Method of Probability Density Quantiles(pdQ), Method of TL-moments (TL), Percentile Matching Method (PM,) Numerical Maximum Likelihood (ML), Maximum Product of Spacings (MPS), Titterington's Method (TM), Starship Method (SM), and Method of Distributional Least Absolutes (DLA). These values are calculated for 500 fitting results for actual parameters $\lambda_1=0$, $\lambda_2=1$, $\lambda_3=2$ and $\lambda_4=0.5$  }
\label{tab3}
\hspace*{-0.5cm}
\begin{adjustbox}{max width=1.3\textwidth}
\begin{tabular}{ccccccccccc}
\toprule
$n$ & Est & pdQ & TL & PM & ML & MPS & TM & SM & DLA\\
\midrule
100 & $\lambda_1$ & 0.020  (0.026) & 0.019  (0.050) & 0.025  (0.057) & 0.024  (0.112) & 0.016*  (0.041) & 0.016*  (0.043) & 0.019  (0.042) & 0.020  (0.032)\\
 & $\lambda_2$ & 0.159  (0.221) & 0.151  (0.206) & 0.448  (0.381) & 0.162  (0.309) & 0.099  (0.158) & 0.096*  (0.150) & 0.184  (0.255) & 0.115  (0.152)\\
 & $\lambda_3$ & 0.396*  (0.252) & 0.446  (0.366) & 0.631  (0.438) & 0.560  (0.645) & 0.527  (0.196) & 0.512  (0.176) & 0.538  (0.323) & 0.545  (0.170)\\
 & $\lambda_4$ & 0.065  (0.133) & 0.035  (0.083) & 0.146  (0.145) & 0.032  (0.097) & 0.028  (0.065) & 0.025*  (0.043) & 0.050  (0.127) & 0.032  (0.054)\\

\midrule
250 & $\lambda_1$ & 0.011*  (0.028) & 0.011*  (0.048) & 0.013  (0.047) & 0.018  (0.112) & 0.012  (0.032) & 0.013  (0.029) & 0.011*  (0.041) & 0.013  (0.013)\\
 & $\lambda_2$ & 0.061  (0.139) & 0.072  (0.151) & 0.218  (0.267) & 0.105  (0.265) & 0.037  (0.090) & 0.034*  (0.075) & 0.081  (0.171) & 0.047  (0.075)\\
 & $\lambda_3$ & 0.324*  (0.235) & 0.336  (0.335) & 0.485  (0.383) & 0.523  (0.640) & 0.526  (0.105) & 0.576  (0.048) & 0.383  (0.296) & 0.553  (0.016)\\
 & $\lambda_4$ & 0.020  (0.088) & 0.018  (0.069) & 0.077  (0.132) & 0.017  (0.090) & 0.009  (0.037) & 0.007*  (0.021) & 0.022  (0.090) & 0.013  (0.032)\\

\midrule
500 & $\lambda_1$ & 0.008*  (0.022) & 0.008*  (0.039) & 0.009  (0.037) & 0.016  (0.111) & 0.009  (0.026) & 0.009  (0.021) & 0.008*  (0.030) & 0.010  (0.004)\\
 & $\lambda_2$ & 0.041  (0.107) & 0.047  (0.118) & 0.109  (0.191) & 0.080  (0.237) & 0.025  (0.070) & 0.023*  (0.055) & 0.044  (0.119) & 0.029  (0.049)\\
 & $\lambda_3$ & 0.276  (0.195) & 0.264*  (0.275) & 0.348  (0.315) & 0.468  (0.631) & 0.365  (0.100) & 0.400  (0.042) & 0.301  (0.208) & 0.414  (0.021)\\
 & $\lambda_4$ & 0.013  (0.069) & 0.012  (0.057) & 0.042  (0.108) & 0.012  (0.082) & 0.005  (0.030) & 0.004*  (0.017) & 0.011  (0.065) & 0.007  (0.025)\\

\midrule

1000 & $\lambda_1$ & 0.006  (0.015) & 0.006  (0.032) & 0.005*  (0.026) & 0.013  (0.101) & 0.006  (0.034) & 0.006  (0.031) & 0.006  (0.021) & 0.008  (0.006)\\
 & $\lambda_2$ & 0.025  (0.078) & 0.029  (0.092) & 0.058  (0.132) & 0.061  (0.203) & 0.016  (0.067) & 0.015*  (0.058) & 0.025  (0.080) & 0.019  (0.022)\\
 & $\lambda_3$ & 0.243  (0.143) & 0.205*  (0.229) & 0.238  (0.230) & 0.398  (0.576) & 0.221  (0.174) & 0.246  (0.140) & 0.220  (0.142) & 0.376  (0.088)\\
 & $\lambda_4$ & 0.008  (0.054) & 0.007  (0.045) & 0.025  (0.081) & 0.009  (0.071) & 0.003  (0.026) & 0.002*  (0.018) & 0.006  (0.044) & 0.004  (0.013)\\
\bottomrule 
\multicolumn{2}{l}{\textsuperscript{*}{Lowest MSE}}
 \end{tabular}
 \end{adjustbox}
\end{table}

%% file: table4.tex
\begin{table}[h]

  \centering
  
    \caption{This table shows a comparison of the mean standard error and the absolute bias (in brackets) of the GLD parameters represented by $\lambda_1$,$\lambda_2$,$\lambda_3$ and $\lambda_4$ between several fitting methods. The considered methods are Method of Probability Density Quantiles(pdQ), Method of TL-moments (TL), Percentile Matching Method (PM,) Numerical Maximum Likelihood (ML), Maximum Product of Spacings (MPS), Titterington's Method (TM), Starship Method (SM), and Method of Distributional Least Absolutes (DLA). Theses values are calculated for 500 fitting results for actual parameters $\lambda_1=0$, $\lambda_2=1$, $\lambda_3=0.5$ and $\lambda_4=0.6$  }
\label{tab4}
\hspace*{-0.5cm}
\begin{adjustbox}{max width=1.3\textwidth}
\begin{tabular}{ccccccccccc}
\toprule
$n$ & Est & pdQ & TL & PM & ML & MPS & TM & SM & DLA\\
\midrule
100 & $\lambda_1$ & 0.027  (0.011) & 0.020  (0.004) & 0.044  (0.005) & 0.022  (0.016) & 0.018*  (0.012) & 0.018*  (0.011) & 0.020  (0.006) & 0.033  (0.016)\\
 & $\lambda_2$ & 0.052  (0.007) & 0.048  (0.043) & 0.213  (0.052) & 0.049  (0.130) & 0.038  (0.002) & 0.037*  (0.013) & 0.053  (0.061) & 0.071  (0.041)\\
 & $\lambda_3$ & 0.038  (0.064) & 0.029  (0.009) & 0.270  (0.122) & 0.049  (0.142) & 0.024*  (0.011) & 0.026  (0.037) & 0.030  (0.026) & 0.134  (0.109)\\
 & $\lambda_4$ & 0.086  (0.103) & 0.044*  (0.005) & 0.323  (0.137) & 0.066  (0.172) & 0.053  (0.035) & 0.048  (0.058) & 0.045  (0.016) & 0.238  (0.158)\\

\midrule
250 & $\lambda_1$ & 0.010  (0.005) & 0.008  (0.002) & 0.018  (0.009) & 0.007*  (0.002) & 0.007*  (0.000) & 0.007*  (0.001) & 0.008  (0.001) & 0.011  (0.002)\\
 & $\lambda_2$ & 0.015  (0.018) & 0.017  (0.028) & 0.092  (0.006) & 0.012  (0.056) & 0.010  (0.001) & 0.009*  (0.010) & 0.016  (0.030) & 0.029  (0.021)\\
 & $\lambda_3$ & 0.010  (0.039) & 0.010  (0.016) & 0.142  (0.088) & 0.009  (0.053) & 0.005*  (0.001) & 0.006  (0.016) & 0.010  (0.019) & 0.053  (0.049)\\
 & $\lambda_4$ & 0.012  (0.048) & 0.011  (0.018) & 0.177  (0.111) & 0.012  (0.061) & 0.007*  (0.003) & 0.007*  (0.018) & 0.010  (0.018) & 0.065  (0.059)\\

\midrule
500 & $\lambda_1$ & 0.004  (0.003) & 0.003*  (0.001) & 0.007  (0.005) & 0.003*  (0.003) & 0.003*  (0.002) & 0.003*  (0.001) & 0.003*  (0.000) & 0.004  (0.001)\\
 & $\lambda_2$ & 0.008  (0.016) & 0.008  (0.025) & 0.043  (0.011) & 0.005  (0.026) & 0.004*  (0.007) & 0.004*  (0.001) & 0.007  (0.022) & 0.010  (0.005)\\
 & $\lambda_3$ & 0.005  (0.027) & 0.005  (0.017) & 0.057  (0.049) & 0.003  (0.024) & 0.002*  (0.007) & 0.002*  (0.005) & 0.004  (0.016) & 0.009  (0.014)\\
 & $\lambda_4$ & 0.007  (0.035) & 0.005  (0.019) & 0.080  (0.063) & 0.003  (0.028) & 0.002*  (0.003) & 0.002*  (0.007) & 0.004  (0.015) & 0.012  (0.018)\\

\midrule

1000 & $\lambda_1$ & 0.002  (0.005) & 0.002  (0.003) & 0.004  (0.006) & 0.001*  (0.004) & 0.001*  (0.004) & 0.001*  (0.003) & 0.002  (0.004) & 0.002  (0.005)\\
 & $\lambda_2$ & 0.004  (0.006) & 0.004  (0.019) & 0.022  (0.010) & 0.002*  (0.016) & 0.002*  (0.005) & 0.002*  (0.003) & 0.003  (0.013) & 0.005  (0.003)\\
 & $\lambda_3$ & 0.002  (0.012) & 0.002  (0.017) & 0.024  (0.003) & 0.001*  (0.013) & 0.001*  (0.006) & 0.001*  (0.003) & 0.002  (0.012) & 0.004  (0.005)\\
 & $\lambda_4$ & 0.003  (0.019) & 0.003  (0.015) & 0.030  (0.014) & 0.001*  (0.017) & 0.001*  (0.002) & 0.001*  (0.006) & 0.002  (0.008) & 0.005  (0.011)\\

\bottomrule
\multicolumn{2}{l}{\textsuperscript{*}{Lowest MSE}}
\end{tabular}
\end{adjustbox}
 \end{table}

%% file: table5.tex
\begin{table}[h]
  \centering
  
    \caption{Parameters chosen for the Generalized Lambda Distribution using the pdQ method and quartile matching.}
\label{tab5}
\hspace*{-0.5cm}
\begin{tabular}{lcccc}
\toprule
Distribution & $\lambda_1$ & $\lambda_2$ & $\lambda_3$ & $\lambda_4$ \\
\midrule
Normal (0,1) & 0 & 1.4420 & 0.1469 & 0.1469 \\
Lognormal (0,1) & 0.8038 & 1.8141 & 0.7589 & -0.7082 \\
$\chi^2_5$ & 4.0559 & 0.4977 & 0.5167 & -0.1470 \\
Beta (2,3) & 0.3770 & 5.3836 & 0.4958 & 0.2637 \\
\bottomrule
\end{tabular}
 \end{table}

%% file: table6.tex
\begin{table}
\footnotesize
 \centering
\caption{This table shows the coverage probability (cp) and mean width ($\Bar{\omega}$) of bootstrap confidence intervals for the location ($\lambda_1$) of the closest GLD distribution at nominal level 95\%. These values are obtained using 500 bootstrap re samples from and 500 iterations}
\label{tab6}
\begin{tabular}{ccccc@{\hskip 0.3in}cc@{\hskip 0.3in}cc}
\toprule
\textit{n} & \textit{F} && \multicolumn{2}{c}{pdQ} & \multicolumn{2}{c}{TM} & \multicolumn{2}{c}{PM}\\
 &  & &   perc & bca & perc & bca & perc & bca\\
\midrule
100 & Normal &  cp  & 0.966 & 0.966 
& 0.964 & 0.960 
& 0.966 & 0.942\\
& &$\Bar{\omega}$  & 0.560 & 0.561 
& 0.489 & 0.490 
& 0.698 & 0.701\\

\cmidrule{2-9}

 &$\chi^2_5$ &  cp  & 0.974 & 0.968 
& 0.946 & 0.956 
& 0.978 & 0.918\\
& &$\Bar{\omega}$ & 1.735 & 1.734 
& 1.500 & 1.453 
& 2.072 & 2.035\\

\cmidrule{2-9}

& Lognormal &  cp & 0.988 & 0.960 
& 0.950 & 0.950 
& 0.968 & 0.882\\
&  & $\Bar{\omega}$& 0.578 & 0.572 
& 0.493 & 0.471 
& 0.682 & 0.660\\

\cmidrule{2-9}

 & Beta &  cp & 0.968 & 0.964 
& 0.948 & 0.960 
& 0.956 & 0.940\\
& &$\Bar{\omega}$ & 0.142 & 0.142 
& 0.123 & 0.121 
& 0.174 & 0.173\\

\midrule

250 & Normal &  cp & 0.948 & 0.948 
& 0.960 & 0.966 
& 0.942 & 0.948\\
& &$\Bar{\omega}$ & 0.341 & 0.342 
& 0.290 & 0.291 
& 0.394 & 0.397\\

\cmidrule{2-9}

 & $\chi^2_5$ & cp & 0.974 & 0.966 
& 0.978 & 0.974 
& 0.960 & 0.944\\
 &  & $\Bar{\omega}$ & 1.010 & 1.011 
& 0.822 & 0.824 
& 1.362 & 1.350\\
 
\cmidrule{2-9}

& Lognormal &  cp & 0.972 & 0.964 
& 0.956 & 0.950 
& 0.962 & 0.928\\
&  & $\Bar{\omega}$ & 0.347 & 0.363 
& 0.282 & 0.270 
& 0.498 & 0.484\\

\cmidrule{2-9}

 & Beta &  cp & 0.950 & 0.934 
& 0.934 & 0.940 
& 0.944 & 0.942\\
&  & $\Bar{\omega}$ & 0.083 & 0.083 
& 0.067 & 0.067 
& 0.107 & 0.107\\
\bottomrule
\end{tabular}
\end{table}

%% file: table7.tex
\begin{table}
\footnotesize
 \centering
\caption{This table shows the coverage probability (cp) and mean width ($\Bar{\omega}$) of bootstrap confidence intervals for the $\lambda_3-\lambda_4$ of the closest GLD distribution at nominal level 95\%. These values are obtained using 1000 and 2000 bootstrap re samples subsequently for n=500 and n=1000 and 500 iterations}
\label{tab7}
\begin{tabular}{ccccc@{\hskip 0.35in}cc}
\toprule
\textit{n} & \textit{F} && \multicolumn{2}{c}{pdQ} & \multicolumn{2}{c}{PM}\\
 &  & & perc & bca &  perc & bca \\
 
 \midrule
500 & Normal & cp  & 0.972 & 0.964 
 & 0.952 & 0.918\\
& &$\Bar{\omega}$ & 0.251 & 0.253 
& 0.457 & 0.465\\

\cmidrule{2-7}

 &$\chi^2_5$ & cp & 0.950 & 0.962 
& 0.956 & 0.934\\
& &$\Bar{\omega}$ & 0.311 & 0.297 
& 0.743 & 0.747\\

\cmidrule{2-7}

& Lognormal & cp & 0.876 & 0.942 
& 0.978 & 0.918\\
& &$\Bar{\omega}$ & 0.560 & 0.477 
& 1.541 & 1.450\\

\cmidrule{2-7}

& Beta & cp & 0.978 & 0.972 
&0.972 & 0.952\\
& &$\Bar{\omega}$ & 0.295 & 0.300 
& 0.646 & 0.650\\

\midrule

1000 & Normal & cp & 0.940 & 0.942 
& 0.958 & 0.934\\
& &$\Bar{\omega}$ & 0.167 & 0.167 
& 0.305 & 0.304\\

\cmidrule{2-7}

&$\chi^2_5$ & cp & 0.944 & 0.956
& 0.964 & 0.940\\
& &$\Bar{\omega}$ & 0.194 & 0.193 
& 0.415 & 0.415\\

\cmidrule{2-7}

& Lognormal & cp & 0.894 & 0.934 
& 0.956 & 0.932\\
& &$\Bar{\omega}$ & 0.304 & 0.293 
& 1.012 & 0.996\\

\cmidrule{2-7}

& Beta & cp & 0.976 & 0.968 
& 0.948 & 0.940\\
& &$\Bar{\omega}$ & 0.172 & 0.172 
& 0.383 & 0.383\\

\bottomrule
\end{tabular}
\end{table}

%% file: table8.tex
\begin{table}[h!t]
 \centering
\caption{The KS test statistic for Spanish household data}
\label{tab8}
\begin{tabular}{ccccccccc}
\toprule
pdQ & TL & PM & ML & MPS & TM & SM & DLA\\
\midrule
0.0069 & 0.0043 & 0.0091 & 0.0326 & 0.0326 & 0.0326 & 0.0183 & 0.0155\\
\bottomrule
\end{tabular}
\end{table}

%% file: table9.tex
\begin{table}
 \centering
\caption{Running times of each method for spanish household data}
\label{tab9}
\begin{tabular}{cccccc}
\toprule
Method & Elapsed time (seconds) & Relative to pdQ \\
\midrule
PM & 0.02 & 0.05\\
pdQ & 0.41 & 1\\
TL & 2.13 & 5.20\\
ML & 11.62 & 28.34\\
TM & 17.19 & 41.93\\
MPS & 17.22 & 42\\
DLA & 21.03 & 51.29\\
SM & 37.39 & 91.20\\
\bottomrule
\end{tabular}
\end{table}

%% file: table10.tex
\begin{table}
 \centering
\caption{The KS test statistic and p-value for twin study data}
\label{tab10}
\begin{tabular}{lrrrrrrrr}
\toprule
  & pdQ & TL & PM & ML & MPS & TM & SM & DLA\\
\midrule
D & 0.0465 & 0.0447 & 0.0583 & 0.0519 & 0.0487 & 0.0487 & 0.0417 & 0.0494\\
p-value & 0.9501 & 0.9657 & 0.7973 & 0.8993 & 0.9344 & 0.9284 & 0.9791 & 0.9236\\
\bottomrule
\end{tabular}
\end{table}

%% file: table11.tex
\begin{table}
 \centering
\caption{Point and interval estimates for the difference of location for earnings of males and females from 1992 to 1998 in US}
\label{tab11}
\begin{tabular}{ccccc}
\toprule
& & pdQ & TM & PM \\
\midrule
Est.& & 2.276 & 2.282 & 2.288 \\
\midrule
Percentile & CI &  (2.026, 2.674) & (2.011, 2.544) & (1.989, 2.736)\\
& Time(mins) & 0.945 & 213  & 0.405\\
\midrule
BCa & CI & ( 1.968,  2.604 ) & (2.021, 2.545) & (1.929,  2.631)\\
& Time(mins) & 59.17 & 6451.2 & 45.03\\
\bottomrule
\end{tabular}
\end{table}

%% file: table12.tex
\begin{table}
 \centering
\caption{Point and interval estimates for the difference of l3 and l4 for earnings of males from 1992 to 1998 in US}
\label{tab12}
\begin{tabular}{cccc}
\toprule
& pdQ & TM & PM \\
\midrule
Est.& 0.383 & 0.351 & 0.417\\
CI-Perc  &(0.365, 0.398) & (0.328, 0.370) & (0.385, 0.459)\\
CI-BCa   &(0.345, 0.418) & (0.329, 0.373) & (0.353, 0.496)\\
\bottomrule
\end{tabular}
\end{table}

%% file: table13.tex
\begin{table}
 \centering
\caption{Point and interval estimates for the difference of l3 and l4 for earnings of females from 1992 to 1998 in US}
\label{tab13}
\begin{tabular}{cccc}
\toprule
& pdQ & TM & PM \\
\midrule
Est.& 0.486 & 0.410  & 0.524\\
CI-Perc  &(0.452, 0.529) & (0.385, 0.436) & (0.455, 0.618)\\
CI-BCa   &(0.447, 0.524) & (0.385, 0.437) & (0.440, 0.602)\\
\bottomrule
\end{tabular}
\end{table}

%% file: main.bbl
\begin{thebibliography}{}

\bibitem[\protect\citename{Alonso-Colmenares {\em et~al.\ }\relax,
  }1994]{alonso1994encuesta}
{\sc Alonso-Colmenares, M.D., Ar{\'e}valo, R., Lara, A., \& Ruiz-Castillo, J.}
  1994.
\newblock {\em La encuesta de presupuestos familiares de 1980-81}.
\newblock {Universidad Carlos III}.

\bibitem[\protect\citename{Asquith, }2007]{asquith2007moments}
{\sc Asquith, W.H.} 2007.
\newblock L-moments and tl-moments of the generalized lambda distribution.
\newblock {\em {Computational Statistics \& Data Analysis}}, {\bf 51}(9),
  4484--4496.

\bibitem[\protect\citename{Chalabi {\em et~al.\ }\relax,
  }2012]{chalabi2012flexible}
{\sc Chalabi, Y., Scott, D.J., \& Wuertz, D.} 2012.
\newblock Flexible distribution modeling with the generalized lambda
  distribution.

\bibitem[\protect\citename{Cheng \& Amin, }1983]{cheng1983estimating}
{\sc Cheng, R.C.H., \& Amin, N.A.K.} 1983.
\newblock Estimating parameters in continuous univariate distributions with a
  shifted origin.
\newblock {\em {Journal of the Royal Statistical Society: Series B
  (Methodological)}}, {\bf 45}(3), 394--403.

\bibitem[\protect\citename{Corlu \& Meterelliyoz, }2016]{corlu2016estimating}
{\sc Corlu, C.G., \& Meterelliyoz, M.} 2016.
\newblock Estimating the parameters of the generalized lambda distribution:
  Which method performs best?
\newblock {\em {Communications in Statistics-Simulation and Computation}}, {\bf
  45}(7), 2276--2296.

\bibitem[\protect\citename{Croissant, }2016]{croissant2016Ecdat}
{\sc Croissant, Y.} 2016.
\newblock {\em {Ecdat: Data Sets for Econometrics}}.
\newblock R package version 5.1.6.

\bibitem[\protect\citename{Dean, }2013]{dean2013improved}
{\sc Dean, B.} 2013.
\newblock {\em Improved estimation and regression techniques with the
  generalised lambda distribution}.
\newblock Ph.D. thesis, {}University of Newcastle.

\bibitem[\protect\citename{Dedduwakumara {\em et~al.\ }\relax,
  }2019]{doi:10.1080/25742558.2019.1602929}
{\sc Dedduwakumara, D.S., Prendergast, L.A., \& Staudte, R.G.} 2019.
\newblock A simple and efficient method for finding the closest generalized
  lambda distribution to a specific model.
\newblock {\em {Cogent Mathematics \& Statistics}}, {\bf 6}(1), 1602929.

\bibitem[\protect\citename{Efron, }1987]{efron1987better}
{\sc Efron, B.} 1987.
\newblock Better bootstrap confidence intervals.
\newblock {\em {Journal of the American statistical Association}}, {\bf
  82}(397), 171--185.

\bibitem[\protect\citename{Efron \& Tibshirani, }1994]{efron1994introduction}
{\sc Efron, B., \& Tibshirani, R.J.} 1994.
\newblock {\em An introduction to the bootstrap}.
\newblock {CRC press}.

\bibitem[\protect\citename{Elamir \& Seheult, }2003]{elamir2003trimmed}
{\sc Elamir, E.H., \& Seheult, A.H.} 2003.
\newblock Trimmed l-moments.
\newblock {\em {Computational Statistics \& Data Analysis}}, {\bf 43}(3),
  299--314.

\bibitem[\protect\citename{Falk, }1986]{falk1986estimation}
{\sc Falk, M.} 1986.
\newblock On the estimation of the quantile density function.
\newblock {\em {Statistics \& Probability Letters}}, {\bf 4}(2), 69--73.

\bibitem[\protect\citename{Freimer {\em et~al.\ }\relax, }1988]{freimer1988}
{\sc Freimer, M., Kollia, G., Mudholkar, G.S., \& Lin, C.T.} 1988.
\newblock A study of the generalized {T}ukey lambda family.
\newblock {\em {Communications in Statistics-Theory and Methods}}, {\bf 17},
  3547--3567.

\bibitem[\protect\citename{Jones, }1992]{jones1992estimating}
{\sc Jones, M.C.} 1992.
\newblock Estimating densities, quantiles, quantile densities and density
  quantiles.
\newblock {\em {Annals of the Institute of Statistical Mathematics}}, {\bf
  44}(4), 721--727.

\bibitem[\protect\citename{Karian \& Dudewicz, }1999]{karian1999fitting}
{\sc Karian, Z.A., \& Dudewicz, E.J.} 1999.
\newblock Fitting the generalized lambda distribution to data: a method based
  on percentiles.
\newblock {\em {Communications in Statistics-Simulation and Computation}}, {\bf
  28}(3), 793--819.

\bibitem[\protect\citename{Karian \& Dudewicz, }2000]{karian2000fitting}
{\sc Karian, Z.A., \& Dudewicz, E.J.} 2000.
\newblock {\em Fitting statistical distributions: the generalized lambda
  distribution and generalized bootstrap methods}.
\newblock {Chapman and Hall/CRC}.

\bibitem[\protect\citename{Karian \& Dudewicz, }2003]{karian2003comparison}
{\sc Karian, Z.A., \& Dudewicz, E.J.} 2003.
\newblock {Comparison of GLD fitting methods: superiority of percentile fits to
  moments in L2 norm}.

\bibitem[\protect\citename{Karvanen {\em et~al.\ }\relax,
  }2002]{karvanen2002adaptive}
{\sc Karvanen, J., Eriksson, J., \& Koivunen, V.} 2002.
\newblock Adaptive score functions for maximum likelihood ica.
\newblock {\em {Journal of VLSI Signal Processing Systems for Signal, Image and
  Video Technology}}, {\bf 32}(1-2), 83--92.

\bibitem[\protect\citename{King {\em et~al.\ }\relax, }2016]{Kingt2016gld}
{\sc King, Robert, Dean, Benjamin, \& Klinke, Sigbert}. 2016.
\newblock {\em gld: Estimation and use of the generalised (tukey) lambda
  distribution}.
\newblock R package version 2.4.1.

\bibitem[\protect\citename{King \& MacGillivray, }1999]{king1999starship}
{\sc King, Robert~A.R., \& MacGillivray, H.L.} 1999.
\newblock A starship estimation method for the generalized lambda
  distributions.
\newblock {\em {Australian and New Zealand Journal of Statistics}}, {\bf 41},
  353--374.

\bibitem[\protect\citename{Owen, }1988]{owen1988starship}
{\sc Owen, D.B.} 1988.
\newblock The {S}tarship.
\newblock {\em {Communications in Statistics-Simulation and Computation}}, {\bf
  17}(2), 315--323.

\bibitem[\protect\citename{Parzen, }1979]{parzen1979}
{\sc Parzen, E.} 1979.
\newblock Nonparametric statistical data modeling.
\newblock {\em {Journal of the American statistical association}}, {\bf 74},
  105--121.

\bibitem[\protect\citename{Pfaff, }2016]{pfaff2016financial}
{\sc Pfaff, B.} 2016.
\newblock {\em {Financial risk modelling and portfolio optimization with R}}.
\newblock {John Wiley \& Sons}.

\bibitem[\protect\citename{Prendergast \& Staudte,
  }2016]{prendergast2016exploiting}
{\sc Prendergast, L.A., \& Staudte, R.G.} 2016.
\newblock Exploiting the quantile optimality ratio in finding confidence
  intervals for quantiles.
\newblock {\em Stat}, {\bf 5}(1), 70--81.

\bibitem[\protect\citename{{R Core Team}, }2017]{R}
{\sc {R Core Team}}. 2017.
\newblock {\em {R: A Language and Environment for Statistical Computing}}.
\newblock R Foundation for Statistical Computing, Vienna, Austria.

\bibitem[\protect\citename{Ramberg \& Schmeiser, }1974]{ramberg1974approximate}
{\sc Ramberg, J.S., \& Schmeiser, B.W.} 1974.
\newblock An approximate method for generating asymmetric random variables.
\newblock {\em Communications of the {ACM}}, {\bf 17}(2), 78--82.

\bibitem[\protect\citename{Ranneby, }1984]{ranneby1984maximum}
{\sc Ranneby, B.} 1984.
\newblock The maximum spacing method. an estimation method related to the
  maximum likelihood method.
\newblock {\em {Scandinavian Journal of Statistics}},  93--112.

\bibitem[\protect\citename{Staudte, }2017]{staudte2017}
{\sc Staudte, R.G.} 2017.
\newblock The shapes of things to come: probability density quantiles.
\newblock {\em {Statistics: A Journal of Theoretical and Applied Statistics}},
  {\bf 51}, 782--800.

\bibitem[\protect\citename{Staudte \& Xia, }2018]{staudte2018}
{\sc Staudte, R.G., \& Xia, A.} 2018.
\newblock Divergence from, and convergence to, uniformity of probability
  density quantiles.
\newblock {\em Entropy}, {\bf 20}(5).

\bibitem[\protect\citename{Su, }2007]{su2007numerical}
{\sc Su, S.} 2007.
\newblock Numerical maximum log likelihood estimation for generalized lambda
  distributions.
\newblock {\em {Computational Statistics \& Data Analysis}}, {\bf 51}(8),
  3983--3998.

\bibitem[\protect\citename{Su {\em et~al.\ }\relax, }2007]{su2007fitting}
{\sc Su, S., {\em et~al.\ }\relax}. 2007.
\newblock Fitting single and mixture of generalized lambda distributions to
  data via discretized and maximum likelihood methods: Gldex in r.
\newblock {\em Journal of {S}tatistical {S}oftware}, {\bf 21}(9), 1--17.

\bibitem[\protect\citename{Tarsitano, }2005]{tarsitano2005estimation}
{\sc Tarsitano, A.} 2005.
\newblock Estimation of the generalized lambda distribution parameters for
  grouped data.
\newblock {\em {Communications in Statistics—Theory and Methods}}, {\bf
  34}(8), 1689--1709.

\bibitem[\protect\citename{Titterington, }1985]{titterington1985comment}
{\sc Titterington, D.M.} 1985.
\newblock Comment on “estimating parameters in continuous univariate
  distributions”.
\newblock {\em {Journal of the Royal Statistical Society: Series B
  (Methodological)}}, {\bf 47}(1), 115--116.

\bibitem[\protect\citename{Tukey, }1965]{tukey1965}
{\sc Tukey, J.W.} 1965.
\newblock Which part of the sample contains the information?
\newblock {\em {Proceedings of the National Academy of Sciences}}, {\bf 53},
  127--134.

\bibitem[\protect\citename{Wang, }2015]{wang2015bda}
{\sc Wang, B.} 2015.
\newblock {\em bda: Density estimation for grouped data}.
\newblock R package version 5.1.6.

\bibitem[\protect\citename{Welsh, }1988]{welsh1988asymptotically}
{\sc Welsh, A.H.} 1988.
\newblock Asymptotically efficient estimation of the sparsity function at a
  point.
\newblock {\em {Statistics \& Probability Letters}}, {\bf 6}(6), 427--432.

\end{thebibliography}
